\documentclass[aps,prd,notitlepage,nofootinbib,superscriptaddress,preprintnumbers,10pt]{revtex4-1}
\usepackage{amsmath}
\usepackage{amssymb}
\usepackage{graphicx}
\usepackage{hyperref}

\begin{document}

\title{Stable cosmology in chameleon bigravity}

\preprint{IPMU17-0137, YITP-17-106}

\author{Antonio De Felice}
\email{antonio.defelice@yukawa.kyoto-u.ac.jp}
\affiliation{Center for Gravitational Physics, Yukawa Institute for Theoretical Physics, Kyoto University, 606-8502, Kyoto, Japan}

\author{Shinji Mukohyama}
\email{shinji.mukohyama@yukawa.kyoto-u.ac.jp}
\affiliation{Center for Gravitational Physics, Yukawa Institute for Theoretical Physics, Kyoto University, 606-8502, Kyoto, Japan}
\affiliation{Kavli Institute for the Physics and Mathematics of the Universe (WPI), The University of Tokyo Institutes for Advanced Study, The University of Tokyo, Kashiwa, Chiba 277-8583, Japan}
\affiliation{Laboratoire de Math\'ematiques et Physique Th\'eorique (UMR CNRS 7350), Universit\'e Fran\c cois Rabelais, Parc de Grandmont, 37200 Tours, France}

\author{Michele Oliosi}
\email{michele.oliosi@yukawa.kyoto-u.ac.jp}
\affiliation{Center for Gravitational Physics, Yukawa Institute for Theoretical Physics, Kyoto University, 606-8502, Kyoto, Japan}

\author{Yota Watanabe}
\email{yota.watanabe@ipmu.jp}
\affiliation{Kavli Institute for the Physics and Mathematics of the Universe (WPI), The University of Tokyo Institutes for Advanced Study, The University of Tokyo, Kashiwa, Chiba 277-8583, Japan}
\affiliation{Center for Gravitational Physics, Yukawa Institute for Theoretical Physics, Kyoto University, 606-8502, Kyoto, Japan}

\date{\today}

\begin{abstract}
 The recently proposed chameleonic extension of bigravity theory, by including a scalar field dependence in the graviton potential, avoids several fine-tunings found to be necessary in usual massive bigravity. In particular it ensures that the Higuchi bound is satisfied at all scales, that no Vainshtein mechanism is needed to satisfy Solar System experiments, and that the strong coupling scale is always above the scale of cosmological interest all the way up to the early Universe. This paper extends the previous work by presenting a stable example of cosmology in the chameleon bigravity model. We find a set of initial conditions and parameters such that the derived stability conditions on general flat Friedmann background are satisfied at all times. The evolution goes through radiation-dominated, matter-dominated, and de Sitter eras. We argue that the parameter space allowing for such a stable evolution may be large enough to encompass an observationally viable evolution. We also argue that our model satisfies all known constraints due to gravitational wave observations so far and thus can be considered as a unique testing ground of gravitational wave phenomenologies in bimetric theories of gravity. 
\end{abstract}

\maketitle


\section{Introduction}

Bimetric theories are an intensively studied class of massive gravity theories considered as an alternative to general relativity (GR). On one hand they predict new phenomena, such as the graviton oscillation \cite{DeFelice:2013nba, Narikawa:2014fua}. On the other hand, bimetric theories contain both a massless and a massive spin-2 field. It has been nontrivial to construct a consistent theory of massive gravity. The first bimetric model free of the so-called Boulware-Deser ghost was proposed by Hassan and Rosen \cite{Hassan:2011zd}, based on the de Rham--Gabadadze--Tolley (dRGT) ghost-free massive gravity model \cite{deRham:2010kj}. 

The bigravity \cite{Hassan:2011zd}, although allowing for a stable cosmological evolution, still requires an important fine-tuning of its parameters in order to be consistent. On one hand, it has been shown that to accommodate a stable evolution, the mass parameter $m$ (controlling the graviton potential terms) needs to be generically much larger than today's Hubble parameter, i.e.\ $m \gg H_0$ \cite{Comelli:2012db, DeFelice:2014nja}. This condition forbids the graviton mass to account for the accelerated expansion of the Universe today. On the other hand, one needs another fine-tuning for (i) the Vainshtein mechanism \cite{vmeca} to effectively screen extra forces on Solar System scales, for (ii) letting the theory be differentiable from GR by leaving nontrivial phenomenology, while (iii) satisfying the Higuchi bound $m_T > \mathcal{O}(1) H_0$ \cite{higuchi}, where $m_T$ is the mass of the tensor modes (proportional but not equal to $m$). Finally, the strong coupling is encountered at a rather low scale $\Lambda_3=(M_{\rm Pl}m^2)^{1/3}$ easily by going early enough in the history of the Universe, which makes the need for a (partial) UV completion all the more important. 

In response to these practical issues, it has been recently proposed to add a new chameleonlike degree of freedom to the theory \cite{DeFelice:2017oym}. In this model, the constant coefficients appearing in the graviton potential are promoted to be general functions of the new scalar field $\phi$, and matter is coupled to gravity through a $\phi$-dependent effective metric. In this way, the effective graviton mass $m_T$ becomes environment dependent, so that $m_T^2$ scales as the local energy density of matter $\rho$. This mechanism allows us to evade the need for the Vainshtein mechanism to screen the extra gravitational forces on Solar System scales, and lets the theory be viable against strong coupling, Higuchi bound, and instabilities up to the very early Universe. The scalar field also has a high enough mass to be possibly not detectable by fifth-force experiments \cite{DeFelice:2017oym}. A possible cosmological application of the chameleonic extension of bigravity theory has been studied in \cite{Aoki:2017ffl}.

In this work we study further the model presented in Ref.\ \cite{DeFelice:2017oym}. Indeed, notwithstanding the arguments in favor of stability and wider applicability that were given, it is important to study the compatibility of the theory versus the observed cosmic evolution. First, we present a detailed study of the scaling solutions, including the conditions for stability under homogeneous perturbations. Second, we present the stability conditions derived from studying the action that is quadratic in inhomogeneous linear perturbations around a flat Friedmann-Lema{\^i}tre-Robertson-Walker (FLRW) spacetime. Finally, we present a viable set of parameters and initial conditions that upon numerical integration leads to a stable cosmological evolution, including radiation-dominated, matter-dominated, and de Sitter phases.

The text is organized as follows. In Sec.~\ref{sec:review} we review the chameleon bigravity model presented in Ref.\ \cite{DeFelice:2017oym}, defining the action and the background equations obtained from its variation. In Sec.~\ref{sec:hompert} we present the scaling solutions of the model, and their respective stability under homogeneous perturbations. In Sec.~\ref{sec:noinstability} we discuss inhomogeneous linear perturbations of the model, and the derivation of the stability conditions in a general flat FLRW universe. In Sec.~\ref{sec:numerics} we present the numerical integration, as well as the chosen parameters and initial conditions. Finally, we conclude in Sec.~\ref{sec:conclusion} and briefly present future extensions of this work. 

\section{Review of the model}
\label{sec:review}

\subsection{Action}

The chameleon bigravity model is defined by the total action $S_\textrm{tot} = S_\textrm{EH} +  S_m + S_\phi + S_\textrm{mat}$ \cite{DeFelice:2017oym}. In this model, the usual ghost-free bimetric theory 
is supplemented by a scalar field $\phi$, coupled to both metrics via the promotion of the coefficients found in the graviton potential into the functions $\beta_i(\phi)$. The gravitational part of the action is given explicitly by
\begin{align}
S_\textrm{EH} &= \frac{M_g^2}{2}\int R[g]\sqrt{-g} d^4x + \frac{M_f^2}{2}\int R[f]\sqrt{-f} d^4x\,,\\
S_m &= M_g^2 m^2\int \sum_{i=0}^4 \beta_i(\phi)U_i[s]\sqrt{-g} d^4x\,, \\
S_\phi &= - \frac{1}{2} \int g^{\mu\nu}\partial_\mu\phi\partial_\nu\phi\sqrt{-g} d^4x\,,
\end{align}
where $M_g$ and $M_f$ stand for the respective bare Planck masses of the gravitational $g$ and $f$ sectors. We also define $\kappa\equiv M^2_f/M^2_g$ for later convenience. Just as in the usual bigravity case the construction of the potentials $U_i$ relies on powers of the metric square root $s^\alpha_\beta \equiv (\sqrt{g^{-1}f})^\alpha_\beta$ such that $s^\alpha_\gamma s^\gamma_\beta= g^{\alpha\delta}f_{\delta\beta}$. By defining $T_n \equiv \textrm{Tr}[s^n]$, we have
\begin{align}
U_0 &= 1\,,\quad U_1 = T_1\,,\quad U_2 = \frac{1}{2}[T_1^2 - T_2]\,,\nonumber\\
U_3 &= \frac{1}{6}[T_1^3-3T_2T_1 + 2T_3]\,, \nonumber\\
U_4 &= \frac{1}{24}[T_1^4 -  6T_1^2T_2 + 3T_2^2 + 8T_1T_3 - 6T_4]\,.
\end{align}
The potentials $U_0$ and $U_4$ constitute the two cosmological constants of the metric sectors $g$ and $f$, respectively. The terms $\beta_i(\phi)U_i$ also play the role of potentials for the field $\phi$. Finally, to implement the chameleon mechanism, the matter sector is coupled nonminimally to the metric $g_{\mu\nu}$, i.e.\
\begin{equation}
S_\textrm{mat} = \int \mathcal{L}_\textrm{mat}(\psi,\tilde{g}_{\mu\nu}) d^4x\,,
\end{equation}
where $\psi$ stands for the different matter fields, $\tilde{g}_{\mu\nu} = A^2(\phi)g_{\mu\nu}$, and $A(\phi)$ is a universal coupling function. In order to simplify the treatment, we adopt the choice of general functions $A(\phi)$ and $\beta_i(\phi)$, following Ref.\ \cite{DeFelice:2017oym}. We thus set
\begin{align}
A(\phi) &= e^{\beta\phi/M_g}\nonumber\,,\\
\beta_i(\phi) &= -c_i e^{-\lambda\phi/M_g}\label{eq:toymodel}\,,
\end{align}
with $i\in\{0, \cdots, 4\}$. These choices are sufficient to obtain a scaling solution described in Sec.~\ref{sec:hompert}. We will use these specific functions for our numerical work. 

\subsection{Background equations}

In order to study cosmological backgrounds, we choose a flat FLRW ansatz for both metrics, i.e.
\begin{align}
ds^2_g = - dt^2 + a^2(t)\delta_{ij}dx^idx^j\,,\quad ds^2_f = \xi^2(t)\left[-c^2(t)dt^2 + a^2(t)\delta_{ij}dx^idx^j\right]\,.
\end{align} 
Under these assumptions, the computation of the metric square root $s^\mu_\nu$ becomes much simpler. We further define the Hubble parameters associated with each gravitational sector, $H\equiv \dot{a}/a$ and $H_f\equiv (a\xi)\dot{}/(ac\xi^2)$, where the dot stands for a derivative with respect to the cosmic time $t$. On such a FLRW background, the equations of motion become the two Friedmann equations
\begin{align}
3H^2 &= \frac{1}{M_g^2}\left[\rho A^4 + \frac{1}{2}\dot{\phi}^2\right] + m^2 R(\xi,\phi)\,,\label{eq:friedmannphysical}\\
3 H_f^2 &= \frac{m^2}{4\kappa\xi^3}U_{,\xi}(\xi,\phi)\,,\label{eq:friedmannfiducial}
\end{align}
(with $R$ and $U$ defined below) as well as the two dynamical equations
\begin{align}
2\dot{H} &= -\frac{1}{M_g^2}\left[\left(\rho+P\right) A^4 + \dot{\phi}^2\right] + m^2 \xi (c-1) J(\xi,\phi)\,,\label{eq:backgroundeinsteinphysical}\\
2 \dot{H}_f &= m^2\frac{1-c}{\kappa\xi^2} J(\xi,\phi) \label{eq:backgroundeinsteinfiducial} \,,
\end{align}
(with $J$ defined below) and the equation of motion for the chameleon scalar field
\begin{equation}
\ddot{\phi} + 3H\dot{\phi} = - \alpha A^4\left(\rho - 3P\right) + M^2_gm^2Q_{,\phi}(\xi,\phi)\,,\label{eq:backgroundscalar}
\end{equation}
(with $Q$ defined below). In these equations we have used 
\begin{equation}
R \equiv U - \xi U_{,\xi}/4\,,\quad J \equiv R_{,\xi}/3\,,\quad Q \equiv (c-1)R - cU\,,\quad U \equiv - \left(\beta_4\xi^4 + 4\beta_3 \xi^3 + 6 \beta_2\xi^2 + 4 \beta_1\xi + \beta_0\right)\,,
\end{equation}
and $\rho$ and $P$ are, respectively, the total energy density and pressure of the matter fields.  By combining the Friedmann (\ref{eq:friedmannphysical}) and second Einstein (\ref{eq:backgroundeinsteinphysical}) equations, one obtains an algebraic equation for $c$ in terms of other variables, 
\begin{equation}
c = \frac{12 J \left(H\xi + \dot{\xi}\right) }{\xi \left(12 H J + \dot{\phi} U_{,\xi\phi}\right)}\,.\label{eq:fiduciallapse}
\end{equation}

In order to represent perfect fluids in the latter analysis, one can choose, for instance, to use $k$-essence scalar fields,
\begin{equation}
S_{\textrm{mat},\alpha} = \int P_\alpha(X_\alpha) \sqrt{-\tilde g}\,d^4x\,,
\end{equation}
where $X_\alpha \equiv -\frac{1}{2}{\tilde g}^{\mu\nu}\partial_\mu\psi_\alpha\partial_\nu\psi_\alpha$ is the canonical kinetic term for a scalar field $\psi_\alpha$. One can then identify pressure $P_\alpha$, energy density $\rho_\alpha$, and the sound speed squared $c_{s,\alpha}^2$ in the Jordan frame as
\begin{equation}
  P_\alpha \equiv P_a(X_a)\,,\quad \rho_a \equiv 2 P_{a,X_a} X_a - P_a(X_a)\,.
  \quad c_{s,\alpha}^2\equiv \frac{P_{\alpha,X_\alpha}}{2P_{\alpha,X_\alpha X_\alpha}X_\alpha + P_{\alpha,X_\alpha}}.
  \label{eq:pressuredensitysoundspeed}
\end{equation}

\section{Stability condition of each era under homogeneous perturbations}
\label{sec:hompert}

\subsection{Scaling solutions}
\label{sec:scalingsol}
It is possible to find exact and approximate scaling solutions to Eqs.\ (\ref{eq:friedmannphysical})--(\ref{eq:backgroundscalar}). We find that in radiation- and cosmological-constant-dominated eras there exist exact scaling solutions. In the matter-dominated era one can find an exact scaling solution only for $\beta = 0$. When $0 < \beta \ll 1$ this turns into an approximate scaling solution. For a radiation-dominated or de Sitter Universe, on the other hand, the exact scaling solutions persist for any value of $\beta$. 

From the Friedmann equation (\ref{eq:friedmannfiducial}) for $f_{\mu\nu}$, we can show that both $\xi =$ constant and $c=$ constant in any scaling solution. Assuming a power law behavior of the scale factor, all terms in the Friedmann equations (\ref{eq:friedmannphysical}) and (\ref{eq:friedmannfiducial}) should scale as $t^{-2}$. Then one can immediately see from the graviton potential terms that if $\xi$ is constant, then 
\begin{equation}
\frac{\phi}{M_g} = \frac{2}{\lambda}\ln{\frac{t}{t_i}} = \frac{n}{\lambda}N_e\,,
\end{equation}
where we have used the standard scaling of the scale factor $a(t) \sim t^{2/n}$ (with $n = 4$ for radiation domination and $n = 3$ for matter domination, here with $\beta = 0$) and introduced the $e$-folding number $N_e=\ln{(a(t)/a_i)}$. Here, $t_i$ ($>0$) is the initial time and $a_i=a(t=t_i)$. 
Denoting a derivative with respect to the $e$-folding time by a prime, one obtains
\begin{equation}
\frac{\phi'}{M_g} = \frac{\dot{\phi}}{M_g H} = \frac{n}{\lambda}\,.\label{eq:phidotscaling}
\end{equation}
In the case of an exponential increase of the scale factor, i.e.\ in a purely de Sitter or $\Lambda$-dominated universe, this last equation (\ref{eq:phidotscaling}) can be extended with the value $n = 0$, since all background quantities (excepting the scale factor) can be taken as constant. Finally, we also have
\begin{equation}
\frac{H'}{H}=-\frac{n}{2}\,.
\end{equation}
In a radiation-dominated universe (and in de Sitter) the scaling expressions presented above can be shown to satisfy all background equations trivially.

On the other hand, in a matter-dominated universe, once we adopt the choices in Eq.~(\ref{eq:toymodel}), we combine background equations to find the following condition including $\beta$:
\begin{equation}
\beta\left(\lambda^2-\frac{3c}{c+\kappa\xi^2}\right)=0. \label{eqn:beta=0}
\end{equation}
As $c$ and $\kappa$ are positive, this condition with $\beta\ne 0$ can be satisfied only if $\lambda \leq \sqrt{3}$. Since we are interested in the regime $\lambda\gg\beta$ to have $m_T^2\propto\rho$ \cite{DeFelice:2017oym}, the condition (\ref{eqn:beta=0}) implies that there is no exact scaling solution in a matter-dominated era unless $\beta=0$. However, if $\beta$ is not zero but small enough then the system with $\lambda\gg\beta$ exhibits an approximate scaling behavior. Therefore, we impose that $\beta \approx 0$ to allow for an approximate scaling solution.

\subsection{Stability under homogeneous perturbation of the scaling solutions}

For practicality, the chameleon scalar field and the Hubble expansion rate are rendered dimensionless using mass parameters of the theory, i.e.,
\begin{eqnarray}
\varphi \equiv \phi/M_{g}\,,\quad h \equiv \frac{H}{m}\,.\label{eq:dimensionlessphihubble}
\end{eqnarray}
The equations are then written in terms of $\ln{h}$, $\varphi$, $\xi$, and $c$. Homogeneous perturbations of the fields are defined as 
\begin{equation}
\begin{cases}
\ln{h}=\ln{h_0}-\frac n2 N_e+\epsilon h^{(1)}\,, \\
\varphi=\frac{nN_e}{\lambda}(1+\epsilon\varphi^{(1)})\,, \\
\xi={\bar \xi}+\epsilon \xi^{(1)}\,, \\
c=c^{(0)}+\epsilon c^{(1)}\,,
\end{cases}
\end{equation}
where $\epsilon$ is a small expansion parameter, $h_0$ is the initial background value of $h$, and $\bar{\xi}$ and $c^{(0)}$ are the constant values of $\xi$ and $c$, respectively, for the scaling solutions. The background equations are then expanded to first order in $\epsilon$. After using the zeroth order equations of motion to set, for instance, $c_0,\kappa,c_4$, and the initial amount of matter (either radiation or dust) in terms of $c^{(0)}$ and the other background variables, one can solve the linearized equations for the variables $h^{(1)}$, $\xi^{(1)}$, and $c^{(1)}$ in terms of $\varphi^{(1)}$ and its derivatives. 

Upon making these replacements, one finds the dynamics is uniquely determined by a second-order equation for $\varphi^{(1)}$. This can be written as
\begin{equation}
\varphi^{(1)}{}'' + \left(1+\frac{2}{N_e}\right)\varphi^{(1)}{}' + \mathcal{A}_r\varphi^{(1)}=0\,,
\end{equation}
during radiation domination (with general $\beta$), and
\begin{equation}
\varphi^{(1)}{}'' + \left(\frac{3}{2}+\frac{2}{N_e}\right)\varphi^{(1)}{}' + \mathcal{A}_m\varphi^{(1)}=0\,,
\end{equation}
during matter domination (with $\beta=0$), where
\begin{align}
\mathcal{A}_r&=\frac{1}{N_e}+\frac{\left[\bar{c} d_{r1} \lambda ^2+4 h_0^2 \left(\lambda ^2-4\right)\right] \left[-6 \bar{c}^3 d_{r1} d_{r2} \lambda ^2-3 (\bar{c}+4)
	\bar{c}^2 d_{r1}^2 \lambda ^2+32 \left(\bar{c}^2+5 \bar{c}+2\right) d_{r1} h_0^2+64 \bar{c}^2 d_{r2} h_0^2\right]}{2 h_0^2
	\lambda ^2 \left[\bar{c}^3 d_{r1}^2 \left(8-3 \lambda ^2\right)+16 \bar{c}^2 \left(d_{r1}^2+d_{r1} h_0^2+2 d_{r2} h_0^2\right)+8 \bar{c}
	d_{r1} \left(d_{r1}+10 h_0^2\right)+32 d_{r1} h_0^2\right]}\,,\label{eqn:mathcalAr}\\
\mathcal{A}_m&=\frac{3}{2 N_e}+\frac{\left[\bar{c} d_{m1} \lambda ^2+3 h_0^2 \left(\lambda ^2-3\right)\right] \left[-4 \bar{c}^3 d_{m1} d_{m2} \lambda ^2-4 \bar{c}^2 d_{m1}^2
	\lambda ^2+36 \bar{c}^2 d_{m2} h_0^2+9 (7 \bar{c}+3) d_{m1} h_0^2\right]}{2 h_0^2 \lambda ^2 \left[\bar{c}^3 d_{m1}^2 \left(3-2
	\lambda ^2\right)+6 \bar{c}^2 \left(d_{m1}^2+2 d_{m2} h_0^2\right)+3 \bar{c} d_{m1} \left(d_{m1}+7 h_0^2\right)+9 d_{m1}
	h_0^2\right]}\,,\label{eqn:mathcalAm}
\end{align}
with $\bar{c}=c^{(0)}-1$, and 
\begin{align}
d_{i1}&=c_1\bar{\xi}_i+2c_2\bar{\xi}_i^2+c_3\bar{\xi}_i^3\,, \label{eqn:def-di1} \\
d_{i2}&=c_2\bar{\xi}_i^2+c_3\bar{\xi}_i^3\,, \label{eqn:def-di2}
\end{align}
with $i=r,m$. 
To guarantee the stability during radiation and matter dominations, respectively, it is necessary and sufficient that 
\begin{equation}
\mathcal{A}_r>0\qquad{\rm and}\qquad\mathcal{A}_m>0\,.\label{eq:hompertcondition}
\end{equation}
Here, it is understood that $\bar{\xi}_r$ and $\bar{\xi}_m$ in (\ref{eqn:def-di1}) and (\ref{eqn:def-di2}) are the constant values of $\xi$ in radiation- and matter-dominated epochs, respectively.

\section{Stability conditions of perturbations}
\label{sec:noinstability}

One can define the perturbations of the fields with respect to the spatially flat FLRW background as follows. In the Arnowitt-Deser-Misner (ADM) decomposition, the (perturbed) metrics are written as
\begin{equation}
ds_{g}^2 = - \mathcal{N}^2 dt^2 + \gamma_{ij}(\mathcal{N}^i dt + dx^i)(\mathcal{N}^j dt + dx^j)\,,\quad ds_{f}^2 = - \tilde{\mathcal{N}}^2 dt^2 + \tilde{\gamma}_{ij}(\tilde{\mathcal{N}}^i dt + dx^i)(\tilde{\mathcal{N}}^j dt + dx^j)\\
\end{equation}
One can then decompose the lapses, shifts, and 3D metrics separately as
\begin{eqnarray}
&\mathcal{N} = N(1 + \Phi) \,,\quad \mathcal{N}_{i} = N_{i} + {\delta N}_i\,,\quad \gamma_{ij} = a^2 \delta_{ij} + {\delta \gamma}_{ij}\,,\nonumber\\
&\tilde{\mathcal{N}} = \tilde{N}(1+ \tilde{\Phi})\,,\quad \tilde{\mathcal{N}}_{i} = \tilde{N}_i + {\delta \tilde{N}}_{i}\,,\quad\tilde{\gamma}_{ij} = \tilde{a}^2 \delta_{ij} + {\delta\tilde{\gamma}}_{ij}\,,
\end{eqnarray}
where $\Phi$, $\tilde{\Phi}$, ${\delta N}_i$, ${\delta \tilde{N}}_i$, ${\delta\gamma}_{ij}$, and ${\delta\tilde{\gamma}}_{ij}$ are the perturbations. In particular, we are free to choose $N = 1$ by the time reparametrization invariance, and we also have that $N_i = \tilde{N}_i = 0$ in our particular background. One may use other equivalent definitions of the perturbations; for instance, as long as the background equations of motion are taken into account, any definitions that differ only at second order will be equivalent as far as the quadratic action is concerned. Finally, as  perturbations are studied only linearly and on a spatially homogeneous and isotropic background, one can decompose the perturbations of the shifts and 3D metrics into $SO(3)$ scalar, vector, and tensor representations, i.e.
\begin{eqnarray}
&{\delta N}_i = N a (\partial_i B + B_i)\,,\quad{\delta\gamma}_{ij} = a^2\left[2\delta_{ij}\Psi + \left(\partial_i\partial_j - \frac{\delta_{ij}}{3}
\Delta
\right)E + \partial_{(i}E_{j)}+ h_{ij}\right]\,,\nonumber\\
&{\delta \tilde{N}}_i = \tilde{N} \tilde{a} (\partial_i \tilde{B} + \tilde{B}_i)\,,\quad {\delta\tilde{\gamma}}_{ij} = \tilde{a}^2\left[2\delta_{ij}\tilde{\Psi} + \left(\partial_i\partial_j - \frac{\delta_{ij}}{3}
\Delta
\right)\tilde{E} + \partial_{(i}\tilde{E}_{j)}+ \tilde{h}_{ij}\right]\,,
\end{eqnarray}
where $h_{ij}$, $\tilde{h}_{ij}$, $E_i$, $\tilde{E}_i$, $B_i$, $\tilde{B}_i$ obey tracelessness and transversality, i.e.\ $\delta^{ij}h_{ij} = \partial^{i}h_{ij} = \partial^{i}E_i = \partial^{i}B_i = 0$ and $\delta^{ij}\tilde{h}_{ij} = \partial^{i}\tilde{h}_{ij} = \partial^{i}\tilde{E}_i = \partial^{i}\tilde{B}_i = 0$. The Laplacian is defined as $\Delta \equiv \delta^{kl}\partial_k\partial_l$, and we use the notation $\mathcal{O}_{(ij)} \equiv \frac{1}{2}(\mathcal{O}_{ij} + \mathcal{O}_{ji})$ to denote symmetrization of the indices. The latin indices of partial derivatives and perturbations can be raised and lowered with $\delta^{ij}$ and $\delta_{ij}$. The perturbations of the chameleon scalar field and matter fields are
\begin{equation}
\phi = \bar{\phi}+\delta\phi\,,\quad\psi_\alpha = \bar{\psi}_\alpha+\delta\psi_\alpha\,.
\end{equation}

The full action is then expanded to second order in the linear perturbations just defined. In particular the perturbations to the metric square root can be computed along the lines of \cite{Gumrukcuoglu:2011zh}. The treatment is separated into tensor, vector, and scalar sectors. For later use, we choose to represent the matter content of the Universe by two perfect fluids, thus labeled by $\psi_\alpha$, with $\alpha \in \{1,2\}$. 

\subsection{Tensor perturbations}

The quadratic action for tensor perturbations (written in Fourier space) reduces to
\begin{equation}
\mathcal{L}^{(2)}_{\textrm{T}} = \frac{M_g^2 N a^3}{8 
}\delta^{ik}\delta^{jl}\left\{\frac{\dot{h}_{ij}\dot{h}_{kl}}{N^2} - \frac{k^2
}{a^2}h_{ij}h_{kl} + \frac{\kappa  \xi^2}{c}\left[\frac{\dot{\tilde{h}}_{ij}\dot{\tilde{h}}_{kl}}{N^2} - c^2\frac{k^2
}{a^2}\tilde{h}_{ij}\tilde{h}_{kl}\right] - \frac{\kappa \xi^2}{c + \kappa \xi^2} m_\textrm{T}^2 h^{-}_{ij}h^{-}_{kl}\right\}\,,
\end{equation}
where $h^{-}_{ij} = h_{ij}- \tilde{h}_{ij}$, $k^2=\delta^{ij}k_ik_j$, $k_i$ is the comoving momentum of a perturbation mode, and 
\begin{equation}
m_\textrm{T}^2 = \frac{c + \kappa \xi^2}{\kappa \xi^2}m^2\Gamma,\quad \Gamma=\xi J+\frac{c-1}{2}\xi^2J_{,\xi}.
\end{equation}
In obtaining this form, we have used both Friedmann equations. One obtains a simple no-ghost condition from the tensor sector, i.e.
\begin{equation}
c\geq 0\,.
\end{equation}
The squared sound speeds of the tensor modes are $c_{T,1}^2 = 1$ and $c_{T,2}^2 = c^2$ for $h_{ij}$ and $\tilde{h}_{ij}$, respectively. 

Due to the time dependence of the background geometry, the graviton mass cannot be defined without ambiguities of order $\mathcal{O}(H)$ in general. On the other hand, in de Sitter spacetime with $\xi =$ constant and $c = 1$, it is the combinations $h^{-}_{ij}$ and $h^{+}_{ij} = h_{ij}+ \kappa\xi^2\tilde{h}_{ij}$ that are the two eigenmodes of the mass matrix. In such a case, one can simply rewrite the Fourier space action in the form
\begin{equation}
\mathcal{L}^{(2)}_{\textrm{T,dS}} = \frac{N a^3 M_g^2}{8(1+\kappa\xi^2)
}\delta^{ik}\delta^{jl}\left\{\frac{\dot{h}^{+}_{ij}\dot{h}^{+}_{kl}}{N^2} - \frac{
k^2
}{a^2}h^{+}_{ij}h^{+}_{kl} + \kappa\xi^2\left[\frac{\dot{h}^{-}_{ij}\dot{h}^{-}_{kl}}{N^2} - \frac{
k^2
}{a^2}h^{-}_{ij}h^{-}_{kl} - m_\textrm{T}^2 h^{-}_{ij}h^{-}_{kl}\right] \right\}\,.
\end{equation}
In this case $m_\textrm{T}$ is the mass of the massive mode, and	 both graviton sound speeds are equal to unity. 

\subsection{Vector perturbations}

After integrating out two nondynamical vectorial degrees of freedom (e.g.\ $B_i$ and $\tilde{B}_i$), the quadratic action for vector perturbations reduces to (in Fourier space)
\begin{equation}
\mathcal{L}^{(2)}_{\textrm{V}} = \frac{M_g^2Na^3}{8
}\frac{m^2\kappa\xi^2Jk^2\delta^{ij}}{(c+1)\kappa\xi k^2/a^2 + 2m^2(c+\kappa\xi^2)J}
\left[\frac{\dot{E}^{-}_i\dot{E}^{-}_j}{N^2} - c^2_{\textrm{V}}\frac{
k^2
}{a^2}E^{-}_i E^{-}_j - m_\textrm{V}^2 E^{-}_iE^{-}_j\right]\,,
\end{equation}
where $E^{-}_i = E_i- \tilde{E}_i$ is the only propagating (massive) vector mode, and
\begin{eqnarray}
c^2_{\textrm{V}} =\frac{(c+1)\Gamma}{2\xi J}\,,\quad m^2_\textrm{V} = m^2_\textrm{T}\,.
\end{eqnarray}
The associated no-ghost condition in the UV regime is, using $c>0$ and $\xi>0$,
\begin{equation}
J\geq 0\,.\label{eq:noghost_vector}
\end{equation}
The no-gradient-instability condition, $c^2_{V}\geq 0$, implies
\begin{equation}
\Gamma\geq 0\,.
\end{equation}

\subsection{Scalar perturbations}
\label{sec:NGanddispersion}

The study of the quadratic action for the scalar perturbations requires more work than the vector and tensor sectors. Because of the size of the expressions, we do not give here the full Lagrangian. Instead, we give here the no-ghost conditions, which must be satisfied at all times during the numerical integration, and the squared sound speeds of the scalar sector, which must be positive at all times.  

We start by integrating out four nondynamical degrees of freedom that enforce the Hamiltonian and (longitudinal part of) the momentum constraints (i.e., $\Phi$, $\tilde{\Phi}$, $B$, $\tilde{B}$). One can integrate out as well the would-be Boulware-Deser (BD) ghost, which is rendered nondynamical by the particular structure of the graviton potential term. One can further use the remaining gauge freedom to set, for instance, the spatially flat gauge, $\Psi = E = 0$. Eventually, one finds that in addition to the two matter perturbation modes, one has two scalar degrees of freedom, one from the chameleon scalar and the other from the massive graviton. 

In order to find both no-ghost conditions and dispersion relations, we take the subhorizon limit $k \gg aH$. Indeed, we are solely interested in checking the presence or absence of instabilities in the UV, any IR instability being less problematic~\cite{Gumrukcuoglu:2016jbh}. 

\subsubsection{No-ghost conditions}

In the subhorizon limit, the action can be written schematically as
\begin{equation}
\mathcal{L}^{(2)}_{S,\mathrm{s.h.}} = \frac{N a^3}{2}\left[\frac{\dot{\mathcal{Y}}^{\top}}{N} \mathcal{K}\frac{\dot{\mathcal{Y}}}{N} +  \frac{\dot{\mathcal{Y}}^\top}{N}\mathcal{F}\mathcal{Y} - 
\mathcal{Y}^\top\mathcal{F}\frac{\dot{\mathcal{Y}}}{N} - \mathcal{Y}^\top\mathcal{M}\mathcal{Y}\right],
\end{equation}
where $\mathcal{K}^\top=\mathcal{K}$, $\mathcal{F}^\top=-\mathcal{F}$, $\mathcal{M}^\top=\mathcal{M}$ are $4\times4$ real matrices, and $\mathcal{Y}$ is a vector containing the four remaining dynamical scalar perturbations, each of which may or may not be rescaled by a positive constant coefficient. The kinetic matrix $\mathcal{K}$ can then be diagonalized, yielding the eigenvalues
\begin{align}
\kappa_1 &= \frac{a^4 m^2 M^2_g}{8 H \kappa} \left\{ 3m^2\left( H - H \kappa \xi^2 + 2 H_f \kappa \xi^3 \right) J^2 + 2\kappa \xi^2 J \left[3 H_f H \left(2H_f\xi - 3 H\right) - \frac{1}{4}m^2 \dot{\phi}U_{,\xi\phi}
\right] \right. \nonumber \\
& \left. + 2 H \kappa \xi^2 \left[3 H_f \xi \left(H - H_f\xi\right) J_{,\xi} - 3H_f \dot{\phi} J_{,\phi} - \frac{1}{16}m^2M^2_g 
U_{,\xi\phi}^2\right] \right\}\,,\\
\kappa_2 &= 1\,,\\
\kappa_3 &= \frac{N^2\left(\rho_1 + P_1\right)}{c^2_{s,1}\dot{\phi}^2_1}\,,\\
\kappa_4 &= \frac{N^2\left(\rho_2 + P_2\right)}{c^2_{s,2}\dot{\phi}^2_2}\,,
\end{align}
up to overall positive constant coefficients. Because of some field redefinition used to diagonalize the kinetic matrix, the indices in $\kappa_i$ are arbitrary, but roughly correspond to, respectively, the scalar graviton, the chameleon field, and both matter perturbations. While $\kappa_2\geq 0$ is trivial and $\kappa_3,\kappa_4\geq 0$ translate into the null-energy conditions on matter fields, i.e., $\rho_\alpha + P_\alpha \geq 0$ (where $\alpha$ is an index designing a specific matter field), $\kappa_1\geq 0$ yields a nontrivial no-ghost condition which will be checked at all times during the numerical integration. We also want to monitor the scalar sound speeds squared, which are read off from the dispersion relations in the subhorizon limit. 

\subsubsection{Scalar sound speeds}

The scalar sound speed for high frequency modes can be found by studying the dispersion relation in the subhorizon limit. Two modes propagate with the usual squared sound speeds $c^2_{s,\alpha}$ of perfect fluids and can thus be identified with the matter modes. The product and the sum of the two remaining scalar sound speeds squared, $c^2_{s,1}$, $c^2_{s,2}$, are given by
\begin{equation}
c^2_{s,1}c^2_{s,2}=\frac{\Sigma_1}{\Sigma}\,,
\end{equation}
and
\begin{equation}
c^2_{s,1}+c^2_{s,2}=\frac{\Sigma_1 + \Sigma_2}{\Sigma} + 1\,,
\end{equation}
where
\begin{align}
\Sigma_1 = &\kappa  \xi  H J \left[-16 M_g^2 \dot{\phi}  \left(J_{,\phi} \left\{(6 c+2) H-(5 c+2) \xi  H_f\right\}-\xi  J_{,\xi\phi} \left\{(c+1) \xi  H_f-2 H\right\}\right)\right.\nonumber\\
&\left.+8 \dot{\phi} {}^2 \left(2 M_g^2 J_{,\phi\phi}+\xi  J_{,\xi}\right)+16 A(\phi )^3 M_g^2 A'(\phi ) J_{,\phi} (3 (P+\rho)-4 \rho)+8 \xi  A(\phi )^4 (P + \rho) J_{,\xi}\right.\nonumber\\
&\left.+M_g^2 \left(\xi  \left\{c^2 m^2 M_g^2 U_{,\xi\phi}^2+16 \xi  J_{,\xi\xi} \left(H-\xi  H_f\right) \left(H-c \xi  H_f\right)\right.\right.\right.\\
&\left.\left.\left.+16 J_{,\xi} \left[c \left(-9 \xi  H_f H+5 \xi ^2 H_f^2+6 H^2\right)+2 \xi  H_f \left(2 \xi  H_f-3 H\right)\right]\right\}\right.\right.\nonumber\\
&\left.\left.+16 (c-1)^2 m^2 \xi  M_g^2 J_{,\phi}^2+8 m^2 M_g^2 J_{,\phi} \left\{2Q_{,\phi} - (c-1)c\xi U_{,\xi\phi}\right\}\right)\right]\nonumber\\
&+4 \xi  J^2 \left(-c \kappa  m^2 \xi  M_g^2 U_{,\xi\phi} \dot{\phi} +6 \kappa  H \dot{\phi} {}^2 +6 \kappa  A(\phi )^4 H (P + \rho)\right. \nonumber\\
&\left. +2 H M_g^2 \left\{2 \kappa  \left[-3 (5 c+2) \xi  H_f H+4 (2 c+1) \xi ^2 H_f^2+(9 c-3) H^2\right]+3 (c-1) m^2 \left(\kappa  \xi ^2+1\right) J_{,\xi}\right\}\right)\nonumber\\
&+16 (c+1) \kappa  \xi  H M_g^2 \left(J_{,\phi} \dot{\phi} +\xi  J_{,\xi} \left(\xi  H_f-H\right)\right){}^2 - 24 m^2 M_g^2 J^3 \left(H \left(3 c \kappa  \xi ^2+c-2 \kappa  \xi ^2-2\right)-2 c \kappa  \xi ^3 H_f\right)\,,\\
\Sigma_2 = & H m^2 M_g^4 \kappa \xi^2 (c-1)^2 J \left(U_{,\xi\phi}-4J_{,\phi}\right)^2\,,\\
\Sigma = & -M_g^2 J \left\{\kappa  \xi ^2 H \left[48 H_f J_{,\phi} \dot{\phi} +48 \xi  H_f J_{,\xi} \left(\xi  H_f-H\right)+m^2 M_g^2 U_{,\xi\phi}^2\right]\right. \nonumber\\
& \left. +4 \kappa  \xi ^2 J \left[m^2 U_{,\xi\phi} \dot{\phi} -12 H_f H \left(2 \xi  H_f-3 H\right)\right]-24 m^2 J^2 \left(2 \kappa  \xi ^3 H_f-\kappa  \xi ^2 H+H\right)\right\}\,,
\end{align}
$\rho = \rho_1 + \rho_2$ and $\rho = P_1 + P_2$. If one considers the vector sector no-ghost condition, $J>0$, then $\Sigma_2 < 0$. The scalar sound speeds squared provide new stability conditions, as these need to be real and positive. We thus require that 
\begin{equation}
\frac{\Sigma_1}{\Sigma}>0\,,\quad \frac{\Sigma_1+\Sigma_2}{\Sigma}+1>0\,,\quad \left(\frac{\Sigma_1+\Sigma_2}{\Sigma}+1\right)^2\! -4\frac{\Sigma_1}{\Sigma}>0\,.
\end{equation}

Although we do not give here the analytical expressions for the single squared sound speeds, which would be too large to write, we obtain their numerical value in the next section as part of our numerical example cosmology (see Fig.~\ref{fig:conditions}). The reader may find a discussion on the respective contributions of the chameleon and the scalar graviton to the scalar squared sound speeds in Appendix~\ref{sec:scalarsoundspeedcontribution}.  

\section{Initial conditions and numerical results}

\label{sec:numerics}

\subsection{Set of equations}

Although in principle one can obtain several background equations \textemdash e.g.\ both Friedmann equations, both second Einstein equations, the scalar equation of motion, or the combination Eq.\ (\ref{eq:fiduciallapse}),\textemdash not all the equations will be directly integrated. For instance, this last equation can be used to fix the fiducial function $c$. Similarly, both Friedmann equations can be used to set two parameters or integration constants, as will be shown below. Of the equations cited above, only three will remain to be integrated: both second Einstein equations and the scalar equation of motion. In addition to finding the right set of equations, the choice of adequate initial conditions (ICs) is also essential. In what follows, a subscript $i$ stands for the quantity evaluated at initial time. 

Although in the previous section we were able to derive the results while keeping the functions $\beta_j(\phi)$, $j\in \{0,\cdots,4\}$, and $A(\phi)$ completely general, these need to be specified for the sake of numerical integration. We will thus from now on use the example model defined in (\ref{eq:toymodel}).

Several definitions help render the equations  more practical for the purpose of numerical integration. First of all we consider the equations of motion in $e$-fold time with its initial value being $N_{e,i}=0$. We then define dimensionless variables. We start by using the dimensionless chameleon scalar field, $\varphi$,  and Hubble parameter, $h$, as defined in Eq.\ (\ref{eq:dimensionlessphihubble}).
For the matter energy densities, we split the energy density of the matter fields (in the Jordan frame, for which $a_{{\rm JF}}=A\,a$) as
\begin{equation}
\rho_{{\rm JF}}^{{\rm TOT}} \equiv \frac{R_{ri}}{A^{4}a^{4}}+\frac{R_{di}}{A^{3}a^{3}}+R_{\Lambda i}\,,
\end{equation}
where the subscripts $r$, $d$, and $\Lambda$, indicate the radiation, dust, and cosmological constant, respectively. We then define
\begin{equation}
R_{ri}  = r_{r}\,a_{i}^{4}\,M_{g}^{2}m^{2}\,,\quad R_{di} = r_{d}\,a_{i}^{3}\,M_{g}^{2}m^{2}\,,\quad R_{\Lambda i} = r_{\Lambda}\,M_{g}^{2}m^{2}\,,
\end{equation}
where $r_r$, $r_m$, and $r_\Lambda$ are dimensionless and constant throughout the evolution. Using these definitions, the Friedmann equation for the physical metric becomes
\begin{equation}
3h^{2}=\frac{1}{2}h^{2}{\varphi'}^{2}+e^{-\lambda\varphi}\left(c_{{0}}+3c_{{1}}\xi+3c_{{2}}\xi^{2}+c_{{3}}\xi^{3}\right)+e^{\beta\,\varphi}r_{{d}}e^{-3N_e}+r_{{r}}e^{-4N_e}+e^{4\beta\,\varphi}r_{{l}}\,,
\end{equation}
while the Friedmann equation for the fiducial metric can be written
\begin{equation}
0 = 1 - e^{-\lambda\varphi}\frac{\bar{V}(\xi)}{3h^2\kappa\xi} - \frac{2\lambda\varphi'}{3}\frac{\bar{V}(\xi)}{\bar{J}(\xi)} + \frac{\lambda^2(\varphi')^2}{9}\frac{\bar{V}(\xi)^2}{\bar{J}(\xi)^2} \label{eq:fiducialfriedmann_numerical}
\end{equation}
where as noted previously a prime denotes differentiation with respect to $N$, and we have defined
\begin{equation}
\bar{J}(\xi) = c_1 + 2c_2\xi + c_3\xi^2\,,\quad\bar{V}(\xi) = c_1 + 3c_2\xi + 3c_3\xi^2 + c_4\xi^3\,.
\end{equation}

It is instructive to rewrite the physical Friedmann equation as
\begin{equation}
1 = \Omega^\mathrm{EF}_\Lambda + \Omega^\mathrm{EF}_d + \Omega^\mathrm{EF}_r + \Omega^\mathrm{EF}_k + \Omega^\mathrm{EF}_V\,.\label{eq:friedmann_densityparameters}  
\end{equation}
For this, we have defined the Einstein frame density parameters
\begin{equation}
\Omega^\mathrm{EF}_\Lambda = \frac{e^{4\beta\,\varphi}r_{{l}}}{3h^2}\,,\quad\Omega^\mathrm{EF}_d = \frac{e^{\beta\,\varphi}r_{{d}}e^{-3N_e}}{3h^2}\,,\quad\Omega^\mathrm{EF}_r = \frac{r_{{r}}e^{-4N_e}}{3h^2}\,,\quad \Omega^\mathrm{EF}_k = \frac{(\phi')^2}{6}\,,\quad \Omega^\mathrm{EF}_V = \frac{e^{-\lambda\varphi}(c_{{0}}+3c_{{1}}\xi+3c_{{2}}\xi^{2}+c_{{3}}\xi^{3})}{3h^2}\,.
\end{equation}
The new subscripts $k$ and $V$ indicate contributions from the chameleon kinetic energy and from the graviton potential term, respectively. We can also define the Jordan frame density parameters, using the fact that 
\begin{eqnarray}
H_{{\rm JF}} & \equiv & \frac{1}{a_{{\rm JF}}^{2}}\,\frac{da_{{\rm JF}}}{d\eta}=
\frac{m\,h}{A}\left(\beta\varphi'+1\right)\,,
\end{eqnarray}
where $\eta$ is the conformal time defined by $\eta \equiv \int_0^t \frac{dt'}{a(t')}$. This allows us to write
\begin{equation}
\Omega_{r}^{{\rm JF}}=\frac{R_{ri}}{A^{2}a^{4}}\,\frac{1}{3M_{g}^{2}H_{{\rm JF}}^{2}}=\frac{r_{r}e^{-4N_e}}{3h^{2}(1+\beta\varphi')^{2}}\,.
\end{equation}
In a similar way, 
\begin{equation}
\Omega_{d}^{{\rm JF}} = \frac{r_{d}\,e^{\beta\varphi-3N_e}}{3h^{2}(1+\beta\varphi')^{2}}\,,\quad \Omega_{\Lambda}^{{\rm JF}}  =  \frac{r_{\Lambda}\,e^{4\beta\varphi}}{3h^{2}(1+\beta\varphi')^{2}}\,.
\end{equation}
Therefore we can replace $r_{r},r_{d},r_{\Lambda}$ with either Jordan frame or Einstein frame density parameters, evaluated at initial time, i.e.,
\begin{eqnarray}
r_{r} & = & 3\Omega_{r,i}^{{\rm JF}}h_{i}^{2}(1+\beta\varphi_{i}')^{2} = 3\Omega_{r,i}^{{\rm EF}}h_{i}^{2}\,,\\
r_{d} & = & 3\Omega_{d,i}^{{\rm JF}}h_{i}^{2}(1+\beta\varphi_{i}')^{2} = 3\Omega_{d,i}^{{\rm EF}}h_{i}^{2}\,,\\
r_{\Lambda} & = & 3\Omega_{\Lambda,i}^{{\rm JF}}h_{i}^{2}(1+\beta\varphi_{i}')^{2}= 3\Omega_{\Lambda,i}^{{\rm EF}}h_{i}^{2}\,.
\end{eqnarray}

In terms of the new variables we have that Eq.\ (\ref{eq:fiduciallapse}) can be rewritten as
\begin{equation}
c\,\xi=\frac{3\left(\xi+\xi'\right)\left(\xi^{2}c_{{3}}+2\,\xi\,c_{{2}}+c_{{1}}\right)}{3(c_{{3}}\xi^{2}+2c_{2}\xi+c_{{1}})-(c_{{4}}\xi^{3}+3c_{{3}}\xi^{2}+3c_{{2}}\xi+c_{{1}})\lambda\varphi'}\,,
\end{equation}
which defines $c$ in terms of the other dynamical variables. When using this definition in the fiducial second Einstein equation, this reduces the degree of the equation to 1, with respect to the variable of interest $\xi$.

The set of dynamical equations to be integrated, the two second Einstein equations and the chameleon field equation, can be written as
\begin{equation}
\begin{cases}
h'= h'(h,\xi,\varphi,\varphi')\,,\\
\varphi'' = \varphi''(h,\xi,\varphi,\varphi')\,,\\
\xi' = \xi'(h,\xi,\varphi,\varphi')\,,
\end{cases}
\end{equation}
and, because of the choice of the variables/parameters, they do not
explicitly depend on any scale, e.g., $M_{g}$ or $m$. 

\subsection{Requirements on initial data}

Using a rescaling of the constants one can, without loss of generality, set the values of the ICs $\varphi_{i}$, $\xi_i$, and $h_i$. In detail, this can be done, for example, by (i) redefining $m^2$ to set $\varphi_i = 0$, (ii) redefining the $c_j$ and $M_f$ to set $\xi_i = 1$, and (iii) an additional overall rescaling of the constants $c_j$, which we use to set $h_i = 1$. Once this is done, we only need to give one supplementary IC, i.e.\ $\varphi'_i$. Then the total set of yet needed ICs and parameters is
\begin{gather*}
c_{0},c_{1},c_{2},c_{3},c_{4},\\
r_{r},r_{d},r_{\Lambda},\\
\lambda,\kappa,\beta,\varphi'_{i}.
\end{gather*}
We can use the two Friedmann equations to set two of the parameters (or ICs, in principle). Without loss of generality, we solve them in terms of $c_{0}$ and $\kappa$ (by linear equations).

The initial conditions for the integration are set in a radiation-domination epoch, with the Universe obeying a scaling solution. These initial conditions allow us to recover a cosmology accommodating our Universe. In order to start with a radiation-domination phase, one simply needs to set $0<1-\Omega_{r,i}^{{\rm JF}}\ll 1$. Since we also want to start from a scaling behavior during radiation domination, the remaining ICs and parameters are imposed so that the dynamics of the scale factor and the scalar field satisfy the scaling solution values found in Sec. \ref{sec:scalingsol}, i.e.,
\begin{equation}
h'_{i} \approx -2h_{i}\,,\quad
\varphi'_{i} \approx \varphi'_{{\rm sc}}=\frac{4}{\lambda}\,,\quad
\varphi''_{i} \approx 0\,,\quad \xi'_i\approx 0\,.
\end{equation}

We choose to replace the parameters $c_1$, $c_2$, $c_3$, and $c_4$ with new, more practical and transparent parameters. First, two of the constants can be chosen so that the condition (\ref{eq:noghost_vector}) is always satisfied. This can be done, for example, by letting 
\begin{equation}
c_3 c_1 - c_2^2 = \mathfrak{A}\,,\quad c_1 + 2c_2+ c_3 = \mathfrak{B}\,,
\end{equation}
where both $\mathfrak{A}$ and $\mathfrak{B}$ are positive constants (and new parameters that replace two among $c_1$, $c_2$, and $c_3$), which is sufficient to guarantee that $J > 0$ for any $\xi$. Second, one may use Eq.\ (\ref{eq:fiduciallapse}), while approximating $\xi_i'\approx 0$, to set $c$ at the initial time to a specific value instead of one of the $c_i$'s. Finally, we use the expression of the vector squared sound speed to replace the last parameter.

\subsection{Results}

Based on the previous section, we describe here a set of parameters which allows for an evolution similar to the usual $\Lambda$ cold dark matter models. The values used in our example are
\begin{gather}
c_\mathrm{in} = \frac{101}{100}\,,\quad c_\mathrm{V,in}^2 = 1\,,\quad \mathfrak{A} = 1\,,\quad \mathfrak{B} = 1\,,\nonumber\\
\Omega^\mathrm{EF}_{\Lambda i} = 1\times 10^{-30}\,,\quad \Omega^\mathrm{EF}_{di} = 1\times 10^{-5}\,,\quad\Omega^\mathrm{EF}_{ki} = \frac{3}{200}\,,\quad \Omega^\mathrm{EF}_{Vi} = \frac{1}{200}\,,\nonumber\\
\beta = 1\times 10^{-2}\,,\quad\lambda = \frac{40}{3}\,,\label{eq:numerical_parameters}
\end{gather}
where the subscripts ``in" or ``i" mean the respective initial value. The initial density parameter for radiation, $\Omega^\mathrm{EF}_{ri}$, is directly determined by the Friedmann equation (\ref{eq:friedmann_densityparameters}) at initial time, and all other parameters are fully determined by this set of choices. The only fine-tuned value is $\Omega_{\Lambda i}^\mathrm{EF}$, which we have chosen in order to have $\Omega_{\Lambda}$ of order unity today. In practice, this is the same as the cosmological constant problem today. 

The simple choice of parameters in Eq.\ (\ref{eq:numerical_parameters}) is meant to show that it is possible to obtain a realistic cosmological evolution. It does not recover exactly today's observed values. However, it is possible, by an appropriate choice of constants \textemdash and without fine-tuning anything other than the cosmological constant \textemdash to obtain an evolution fitting more closely to data; e.g.\ one can reproduce today's abundances and other data. This, along with the constraints on the model from today's observational data, will be studied further in a future work.

\begin{figure}[ht]
	\centering
	\includegraphics[width=12cm]{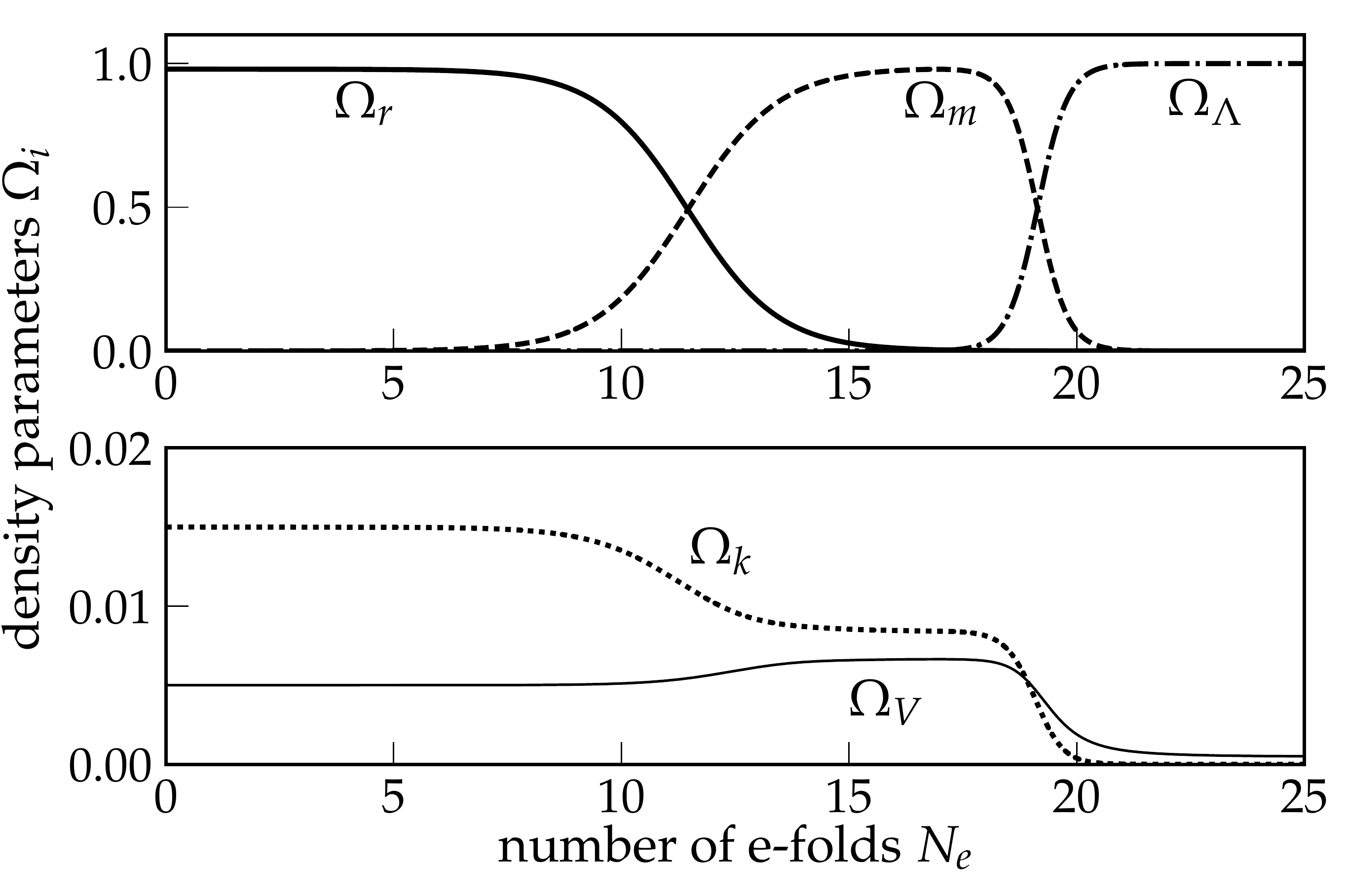}
	\caption{Evolution of the density parameters of the different species versus number of $e$-folds of evolution. The thick solid line stands for radiation, the thick dashed line stands for dust, the dashed-dotted line stands for the cosmological constant, the thin dotted line stands for the scalar field kinetic energy, and the thin solid line stands for the contribution from the graviton potential term.}\label{speciesratio}
\end{figure}

\begin{figure}[ht]
	\centering
	\includegraphics[width=12cm]{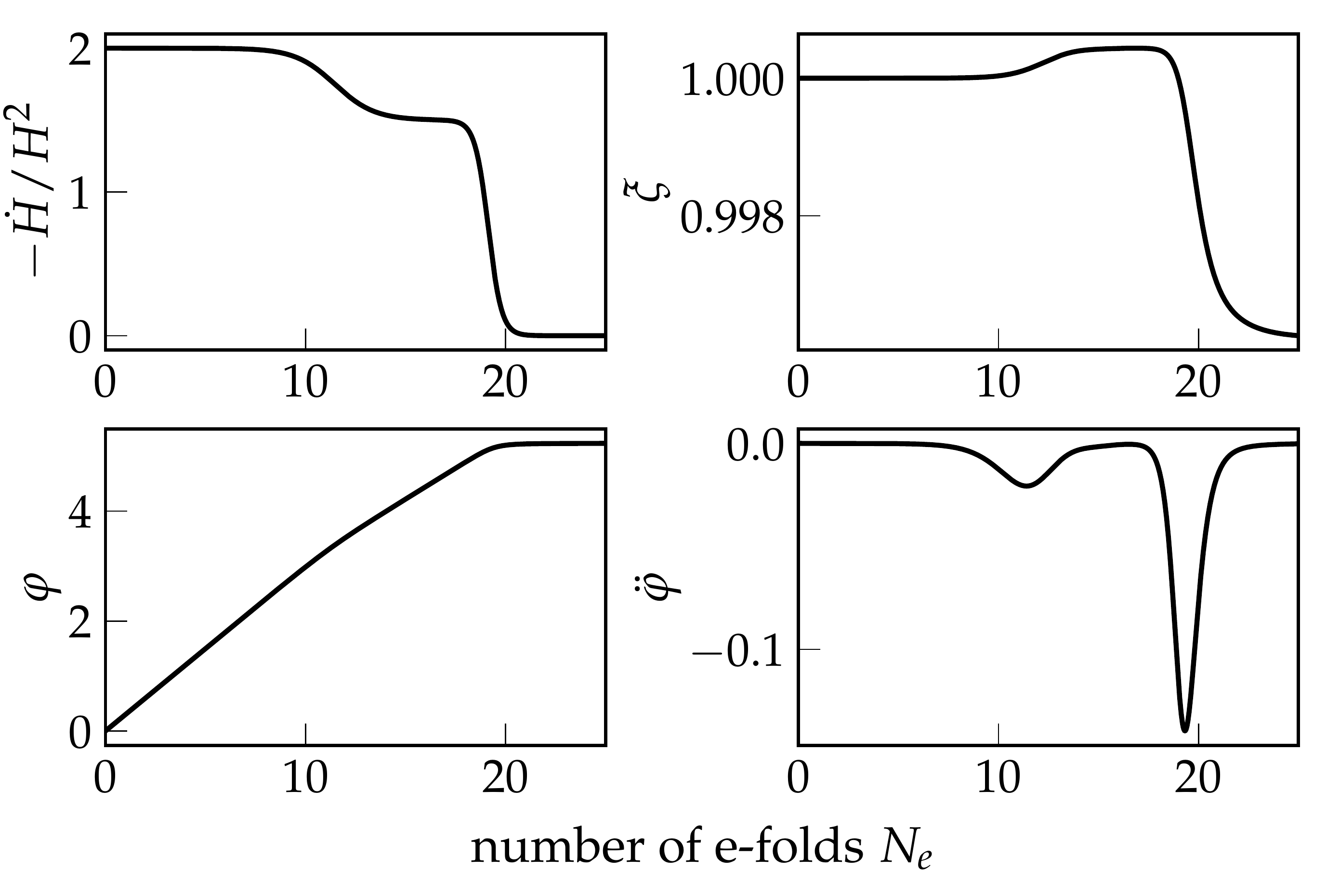}
	\caption{Evolution of the time derivative of the Hubble parameter, the ratio of fiducial and physical scale factors $\xi$, the chameleon field $\varphi$, and its second time derivative. The system starts from a radiation-dominated era (from $N_e = 0$ to roughly $N_e= 12$), then goes through a matter-domination phase (roughly from $N_e = 12$ to roughly $N_e= 19$), and finishes in a de Sitter era.}\label{fig:varevolution}
\end{figure}

For the sake of exposition, we present the evolution\footnote{The number of steps and $e$-fold time range chosen for the integration are  
initial time = 0, final time = 25, and number of steps = 3199.} of the density parameters in Fig.\ \ref{speciesratio}, while the evolution of other relevant variables is presented in Fig.\ \ref{fig:varevolution}. The evolution, starting from a radiation-dominated era, moves on to a matter-dominated era, finally attaining a final de Sitter phase. Given our set of initial density parameters, the system stays $\mathcal{O}(10)$ $e$-folds in each era before settling to a de Sitter epoch (roughly from $0$ to $12$ $e$-folds for radiation domination, from $12$ to $19$ $e$-folds for matter domination, from $19$ to the end for  the de Sitter era). However, by arranging these density parameters, one can achieve very different numbers of $e$-folds spent in each era. 

\begin{figure}[ht]
	\centering
	\includegraphics[width=12cm]{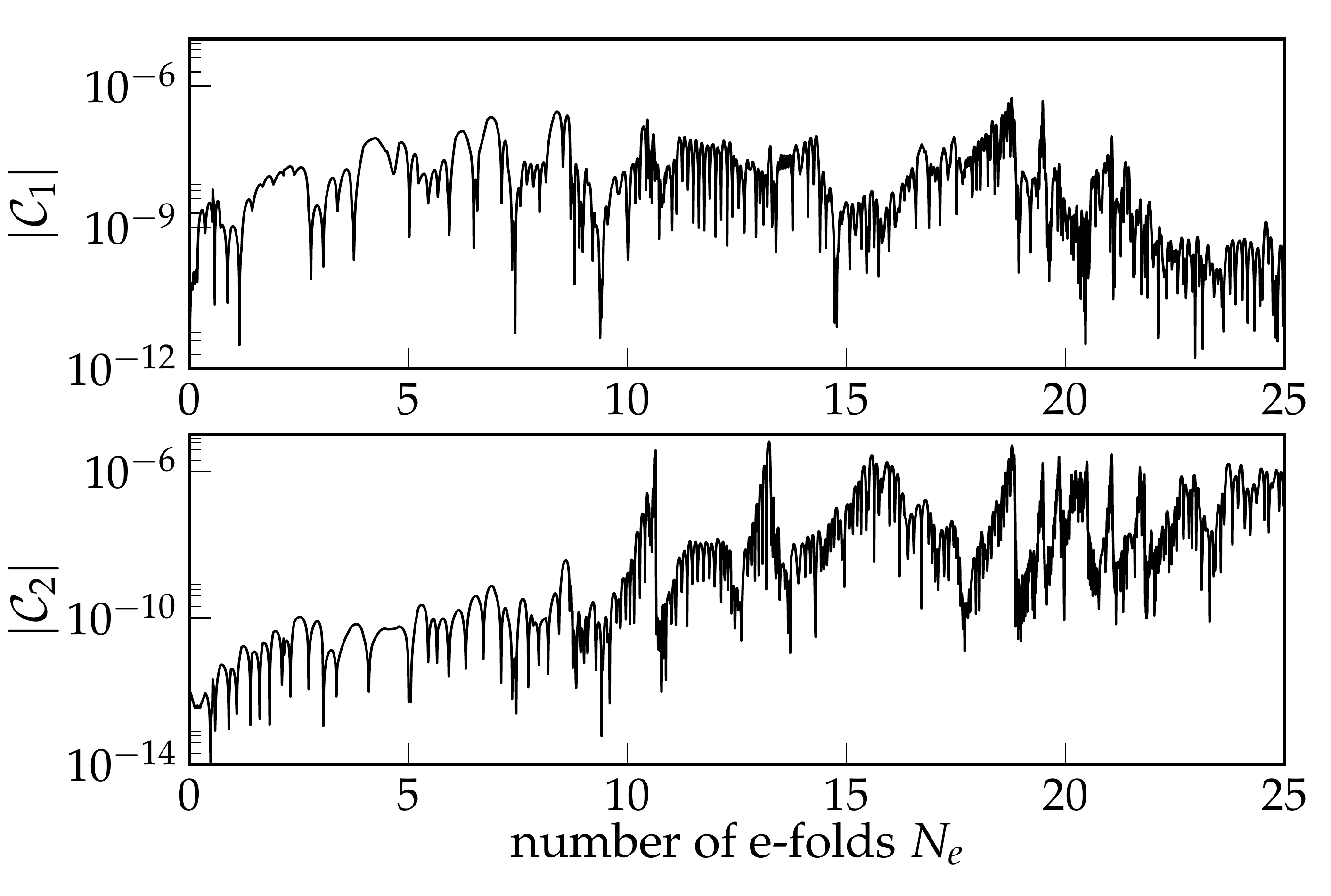}
	\caption{Evolution of the first and second constraints. Constraint damping is efficient during most of the integration time.}\label{fig:constraintevolution}
\end{figure}

In order to have a handle on the precision of the numerical integration, we check all along the evolution to which extent the Friedmann equations are satisfied. For this purpose, one may define, for instance,
\begin{equation}
\mathcal{C}_1 = \frac{1 - \sum_\alpha\Omega^\mathrm{EF}_\alpha}{1 + \sum_\alpha \left|\Omega^\mathrm{EF}_\alpha\right|}\,,\quad \mathcal{C}_2 = \frac{ 1 - e^{-\lambda\varphi}\frac{\bar{V}(\xi)}{3h^2\kappa\xi} - \frac{2\lambda\varphi'}{3}\frac{\bar{V}(\xi)}{\bar{J}(\xi)} + \frac{\lambda^2(\varphi')^2}{9}\frac{\bar{V}(\xi)^2}{\bar{J}(\xi)^2}}{1 + \left|e^{-\lambda\varphi}\frac{\bar{V}(\xi)}{3h^2\kappa\xi}\right| + \left|\frac{2\lambda\varphi'}{3}\frac{\bar{V}(\xi)}{\bar{J}(\xi)}\right| + \left|\frac{\lambda^2(\varphi')^2}{9}\frac{\bar{V}(\xi)^2}{\bar{J}(\xi)^2}\right|}\,,
\end{equation}
inspired by both Friedmann equations (\ref{eq:friedmann_densityparameters}) and (\ref{eq:fiducialfriedmann_numerical}), and where $\alpha$ stands for any of the species, i.e., indices $\{r,d,\Lambda,k,V\}$. The evolution of these two constraints is presented in Fig.\ \ref{fig:constraintevolution}. Both constraints are seen to be satisfied up to order $\mathcal{O}(10^{-5})$. In our implementation, this has been achieved by using constraint damping, i.e., adding the constraint equations into the dynamical equations of motion (after normalizing the constraints by an appropriate factor), in order to damp any unwanted deviation.

\begin{figure}[ht]
	\centering
	\includegraphics[width=12cm]{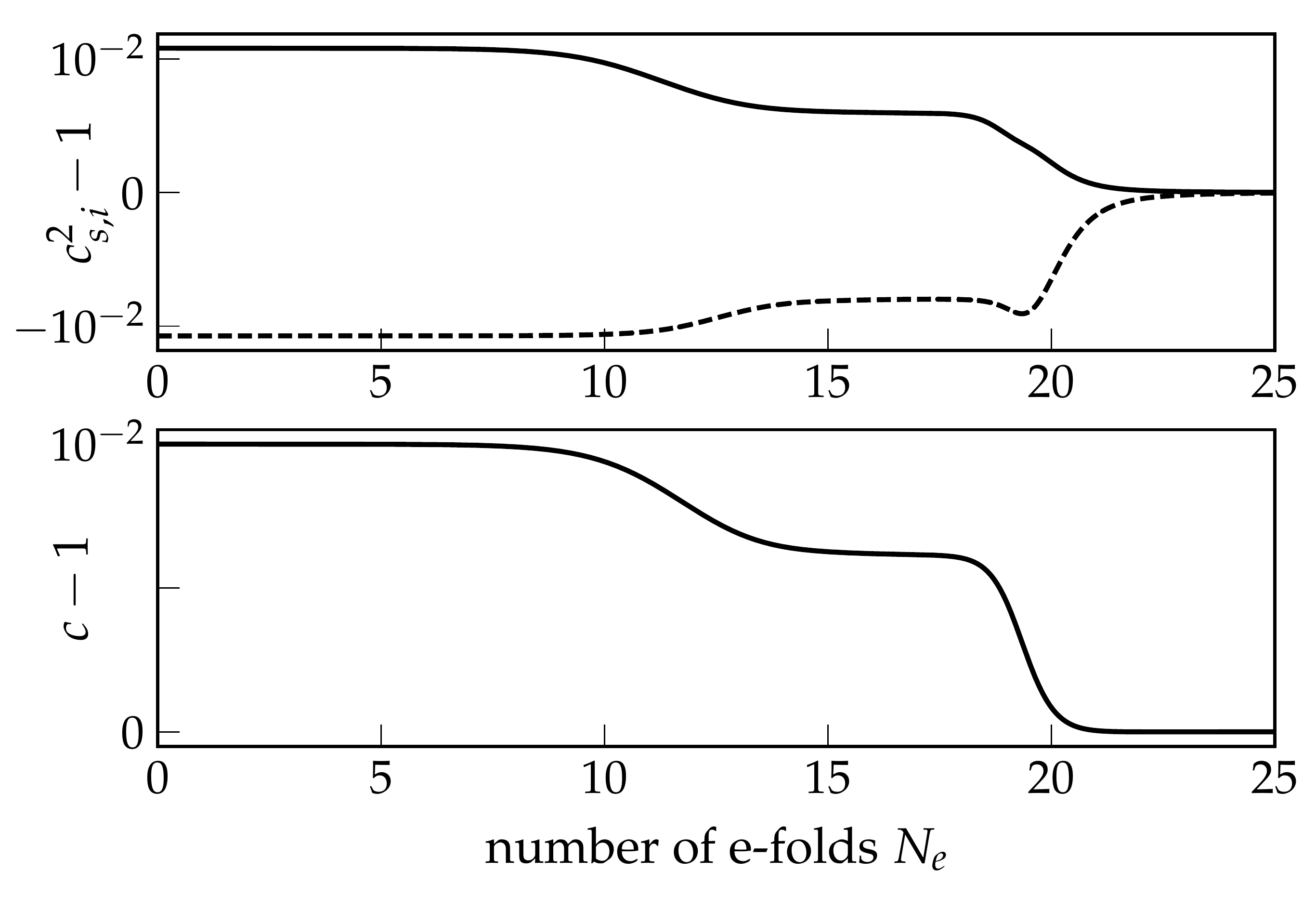}
	\caption{Evolution of some consistency conditions. Here are presented the evolution of the two scalar squared sound speeds ($c^2_{s,i}$ with $i\in\{1,2\}$) and of the fiducial lapse $c$. Both the sound speeds and the fiducial lapse tend rapidly to $1$ in a $\Lambda$-dominated universe. Respective contributions from the scalar graviton and the chameleon scalar field to the scalar sound speeds are discussed in the Appendix.}\label{fig:conditions}
\end{figure}

In addition to the Friedmann equations, we also present in Fig.\ \ref{fig:conditions} the evolution of the sound speeds and the fiducial lapse $c$. Together with the no-ghost conditions, which are found to be satisfied all along the evolution, the positivity of these shows that the background is stable under cosmological perturbations.

\begin{figure}[ht]
	\centering
	\includegraphics[width=12cm]{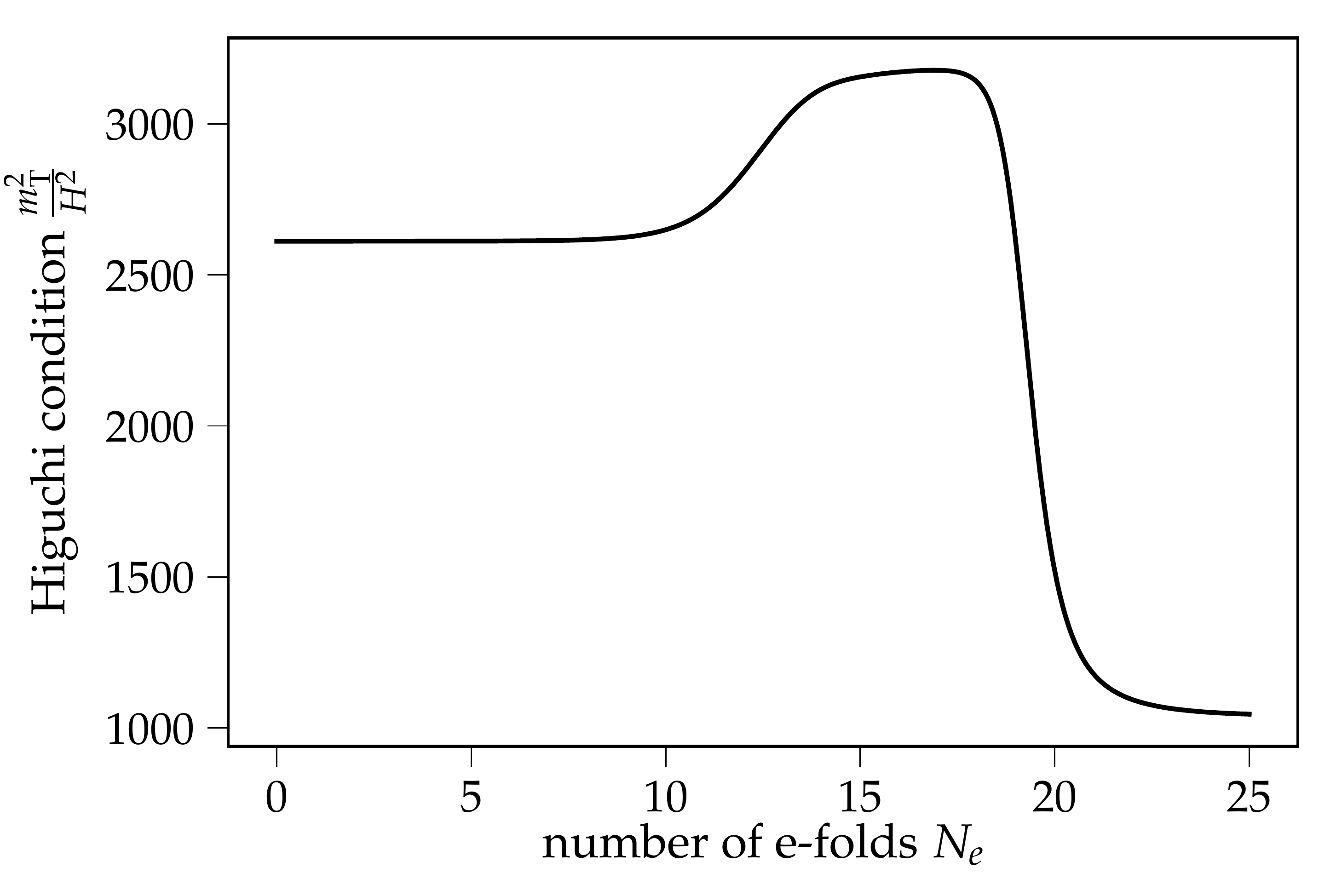}
	\caption{Evolution of the Higuchi condition. The ratio of the tensor mass to the Hubble expansion rate has to be $> \mathcal{O}(1)$ in order for the model to be stable. The condition is thus satisfied all along the evolution.}\label{fig:higuchi}
\end{figure}

Finally, in order to demonstrate the purpose of the new scaling brought by the scalar field dependence in the graviton potential, we plot the Higuchi condition along the evolution in Fig.\ \ref{fig:higuchi}. The generalized Higuchi bound $\frac{m^2_\mathrm{T}}{H^2} > \mathcal{O}(1)$ is seen to be well satisfied during the three eras, and in particular both at late and early times. 

\section{Discussion and conclusions}
\label{sec:conclusion}

Following the recent proposal \cite{DeFelice:2017oym} of an extended massive bigravity theory supplemented by a chameleon scalar field as a means to cure or evade the fine-tunings of the original theory and improve its applicability, we have found it important to study further its validity and implications. For this reason, in this work we have explored the stability conditions of the model and confirmed its intended behavior by integrating numerically the equations of motion.  In particular, we have numerically confirmed that at all times, the Higuchi ghost is never present: indeed, the presence of this ghost represented one of the most serious problems for a viable phenomenology of the original bigravity theory. In our model though, we have here shown that if no Higuchi ghost is present at one scale then the same ghost will not appear during the whole evolution of the Universe including the early epoch. This set of such allowed initial conditions is not of zero measure in general, so that we do not need to fine-tune the parameters of the theory.

The study of the action quadratic in perturbations with respect to a general flat FLRW background leads, in the UV, to no-ghost conditions for the tensor, vector, and scalar sectors. In addition to this, we have found the explicit action for the tensor and vector linear perturbations and for the scalar linear perturbations in the UV. From these the propagation speeds at short scales are easily extracted, thus leading to additional no-instability conditions. It is found as expected that the theory propagates four tensor, two vector, and two scalar degrees of freedom (not including matter degrees of freedom), thus corresponding to the expected massive spin-2, massless spin-2, and chameleon scalar of the theory.

In order to show the typical background time evolution, we have numerically integrated the background equations by using a choice of initial parameters consistent with an initial radiation-dominated era of the Universe. As supplementary input for the initial conditions, we have required that the stability conditions be satisfied and that the parameters of the theory are in the regime of interest for the expected scaling behaviors. The evolution displays an initial radiation-dominated era, followed by matter domination and a de Sitter era. The no-instability conditions are satisfied all along the evolution, and, in our implementation, the constraint equations show a numerical error of order $\mathcal{O}(10^{-5})$ at most. This stable evolution comforts us into arguing that it may be possible to find a region of the parameter space allowing a close match with our cosmological observations.

The recent binary neutron star merger observation, the first gravitational and electromagnetic wave multimessenger detection \cite{GBM:2017lvd}, has allowed us to set stringent bounds on the speed difference between gravitational and electromagnetic waves (see, e.g., \cite{Creminelli:2017sry,Ezquiaga:2017ekz,Baker:2017hug,Sakstein:2017xjx}). Although in our model one of the gravitons propagates with a slightly modified sound speed $c$ (see the lower panel of Fig.~\ref{fig:conditions}), the physical metric remains unaffected and the interactions between the two metrics are suppressed by the smallness of $m^2\beta_i$, $i\in \{1,2,3\}$. This implies that the propagation of gravitational waves in our model is essentially the same as that of photons as far as $m^2\beta_i$ are small enough compared to the typical (squared) energy scales of the gravitational waves produced astrophysically. As a result, the constraint on our model from GW170817 is essentially the same as those from the previous GW observations (e.g., \cite{bib:LIGO})~\cite{Baker:2017hug}. Concretely, the constraint is of the form of an upper bound on the mass of the graviton (which was not improved by GW170817) of $m_T < 1.2 \times 10^{-22}$~eV. While this bound has to and can be satisfied today, the scalar field dependence of the graviton mass in our model allows without problem for a larger mass at early times, rendering the cosmological evolution stable all the time. Therefore, our model can be considered as a unique testing ground of gravitational wave phenomenologies in bimetric theories of gravity. For example, it is intriguing to investigate the possible modification of the waveform of the gravitational wave signal due to the influence of the massive graviton. 

As a clear avenue for future extension, the evolution of cosmological perturbations and an improved understanding of the viable parameter space will be considered in a future work. Furthermore, it may be interesting to study the detailed working of the screening mechanism for the chameleon scalar field and scalar graviton modes.

\appendix

\section{Contribution to scalar sound speeds}
\label{sec:scalarsoundspeedcontribution}

In Fig.~\ref{fig:conditions}, two $c_s^2$'s are plotted. Although each $c_s^2$ is contributed both by the chameleon and by the scalar graviton, the dominant contribution can be determined by the following argument: the $c_s^2$ are determined by
\begin{equation}
\det\left[c_s^2\mathcal{K}_\mathrm{diag}-\mathcal{M}_\mathrm{rot} \right]=0,
\label{eq:cs2}
\end{equation}
where $\mathcal{K}_\mathrm{diag}$ is the kinetic matrix $\mathcal{K}$ made diagonal by some rotation matrix and $\mathcal{M}_\mathrm{rot}$ is the mass matrix $\mathcal{M}$ rotated by the same rotation matrix in the high frequency limit. Those matrices can be written in the form
\begin{equation}
\mathcal{K}_\mathrm{diag}=
 \begin{pmatrix}
     1 & 0 \\ 
     0 & \kappa_1 \\
\end{pmatrix},
\quad
\mathcal{M}_\mathrm{rot}=
\begin{pmatrix}
1 & A \\ 
A & B \\
\end{pmatrix},
\end{equation}
where $A$ and $B$ are some components, since the radiation and dust fluids are decoupled from the chameleon and the scalar graviton in the high frequency limit. On the other hand, Eq.~\eqref{eq:cs2} can be written, introducing eigenvector $(v_1\ v_2)^\top$ and normalizing $v_2$, as
\begin{equation}
\left[c_s^2 I_2-
\begin{pmatrix}
1 & A/\sqrt{\kappa_1} \\ 
A/\sqrt{\kappa_1} & B/\kappa_1 \\
\end{pmatrix}
\right]
\left(
\begin{array}{c}
v_1 \\
v_2
\end{array}
\right)=0,
\end{equation}
in the high frequency limit, where $I_2$ is the $2\times2$ identity matrix. This yields the ratio of $v_1$ to $v_2$,
\begin{equation}
\left|\frac{v_1}{v_2}\right|_\pm=\left|\frac{A/\sqrt{\kappa_1}}{c_\pm^2-1}\right|,
\label{eq:eigenvector}
\end{equation}
where $c_\pm^2$ are the solutions of Eq.~\eqref{eq:cs2}, and whose value can be checked numerically. If Eq.~\eqref{eq:eigenvector} is larger (smaller) than 1, the dominant contribution is the chameleon (the scalar graviton). Our calculation shows that the larger $c_s^2$ in Fig.~\ref{fig:conditions} is dominantly contributed by the chameleon.

Note that one of the ratios $|v_1/v_2|_\pm$ is larger than 1 if the other is smaller than 1 and vice versa, since
\begin{equation}
\left|\frac{v_1}{v_2}\right|_+\left|\frac{v_1}{v_2}\right|_-=\left|\frac{A^2/\kappa_1}{c_+^2c_-^2-(c_+^2+c_-^2)+1}\right|=1,
\end{equation}
which follows from the relation between the solutions $c_\pm^2$ of the quadratic equation \eqref{eq:cs2},
\begin{align}
c_+^2+c_-^2&=1+B/\kappa_1, \\
c_+^2c_-^2&=(B-A^2)/\kappa_1.
\end{align}

\acknowledgments
S.M.\ thanks the Laboratoire de Math\'ematiques et Physique Theorique,
Universit\'e Fran\c{c}ois-Rabelais de Tours for hospitality. 
A.D.F.\ was supported by Japan Society for the Promotion of Science (JSPS) KAKENHI Grant No.\ 16K05348 and No.\ 16H01099. 
The work of S.\,M.\ was supported by JSPS Grants-in-Aid for Scientific Research (KAKENHI) No. JP17H02890 and No. JP17H06359. 
The work of S.\,M.\ and Y.\,W.\ was supported by World Premier International Research Center Initiative (WPI), Ministry of
Education, Culture, Sports, Science and Technology(MEXT), Japan.
The work of Y.\,W.\ was supported by JSPS Grant-in-Aid for Scientific Research No.\,16J06266 and by the Program for Leading Graduate Schools, MEXT, Japan. 
M.\,O.\ acknowledges the support from the Japanese Government (MEXT) Scholarship for Research Students, and thanks C.~Ott and J.~Fedrow for useful explanations on how to plot with Python.

\end{document}